\documentclass[a4paper,12pt]{article}
\usepackage{graphicx}
\usepackage{bm}
\usepackage{amstext,amsmath,amssymb}
\usepackage{mathptmx}
\usepackage[top=3cm,bottom=3cm,left=1.5cm,right=1.5cm,asymmetric]{geometry}
\usepackage{rotating}
\usepackage{hyperref}
\usepackage{cite}
\usepackage{array}
\usepackage{tabularx}
\DeclareMathOperator{\sech}{sech}

\begin{document}

\title{Exactly solvable extended potentials in arbitrary dimensions}
\author{\bf{Nabaratna Bhagawat}i\\
Department of Physics, Gauhati University, Guwahati-781014, India \\
E-mail: nabaratna2008@gmail.com} 
\date{}
\maketitle

\abstract{We apply a simple transformation method to construct a set of new exactly solvable potentials (ESP) which gives rise to bound state solution of $D$-dimensional Schr\"odinger equation. The important property of  such exactly solvable quantum systems is that their normalized eigenfunctions can be written in terms of recently introduced exceptional orthogonal polynomials (EOP).
}

\section{Introduction}
\label{sec1}
The exact solutions of fundamental dynamical equations are important in different areas of physics and chemistry. Researchers always try to find new exact solution of Schr\"odinger equation as it is possible only for a few potentials. Also, exactly solvable potentials are essential for the successful implementation of approximate methods in the study of practical quantum systems.\\
\hspace*{0.3in}Since the early days of quantum mechanics, classical orthogonal polynomials (COP) such as Laguerre, Legendre, Jacobi, Hermite etc. play an important role as the bound state eigenfunctions are expressible in terms of these polynomials. The factorization method~\cite{Schrodinger, Infeld} initiated by Schr\"odinger is the pioneering work in this regard. Thereafter, researchers employed various methods e.g. the point canonical transformation (PCT) method~\cite{Bhattacharjie, Levai1, Levai2}, the supersymmetric (SUSY) method~\cite{Gendenshtein,  Cooper}, the Nikiforov-Uvarov(NU) method~\cite{Nikiforov, Egrifes, Hamzavi}, the Extended Transformation(ET) method~\cite{Ahmed1, Ahmed2}, the asymptotic iteration method (AIM)~\cite{Bayrak}, the new exact quantization rule~\cite{Dong}, the Laplace transforms method~\cite{Chen}, the path integral method~\cite{Grosche} etc. to solve the Schr\"odinger equation analytically and they used COPs to express the bound state eigenfunctions.\\
\hspace*{0.3in}In a recent advancement in the field of mathematics and physics, two new families of exceptional orthogonal polynomials, $\hat{L}_n^{\alpha}(x)$ and $\hat{P}_n^{(\alpha, \beta)}(x)$, $n=1,2,3...$ have been introduced~\cite{Ullate1, Ullate2}. The use of these polynomials in constructing new ESPs~\cite{Quesne1, Quesne2} has already drawn attention in quantum mechanics. In this paper, we use a simple transformation method~\cite{Bhagawati} to construct four new exactly solvable extended potentials using the properties of EOPs. These potentials are ``extended" in the sense that they can be expressed in terms of some well-known exactly solvable potentials and a few additional rational terms.\\
\hspace*{0.3in}The paper is organized as follows. In section~\ref{sec2}, we present a brief overview of the transformation method. The construction of new ESPs using the properties of EOPs is given in detail in section~\ref{sec3}. Conclusions and discussions are covered in section~\ref{sec4}.

\section{The Transformation Method}
\label{sec2}
We consider a second order linear differential equation satisfied by a special function $Q(r)$~\cite{Bhagawati}
\begin{eqnarray}
\label{eq1}
Q^{\prime\prime}(r)+M(r)Q^{\prime}(r)+J(r)Q(r)=0
\end{eqnarray}
where a prime denotes differentiation with respect to its argument. $Q(r)$ will later be identified as one of the orthogonal polynomials.
\\ The transformation method comprises of the following two steps
\begin{equation}
\label{eq2}
r\rightarrow g(r)
\end{equation}
\begin{equation}
\label{eq3}
\psi(r)=f^{-1}(r)Q(g(r))
\end{equation}
We implement the above prescription to equation~(\ref{eq1}) and obtain
\begin{eqnarray}
\label{eq4}
&&\psi^{\prime\prime}(r)+\left(\frac{d}{dr}\ln\frac{f^{2}(r)\exp(\int M(g)dg)}{g^{\prime}(r)}\right)\psi^{\prime}(r)+\nonumber \\
&&\left(\frac{f^{\prime\prime}(r)}{f(r)}-\frac{g^{\prime\prime}(r)}{g^{\prime}(r)}\frac{f^{\prime}(r)}{f(r)}+g^{\prime}(r)M(g)\frac{f^{\prime}(r)}{f(r)}+g^{\prime 2}J(g)\right)\psi(r)=0
\end{eqnarray}
The radial Schrodinger equation in $D$-dimensional Euclidean space is ($\hbar=1=2m$)
\begin{eqnarray}
\label{eq5}
\psi^{\prime\prime}(r)+\frac{(D-1)}{r}\psi^{\prime}(r)+\left(E_{n}-V(r)-\frac{\ell(\ell+D-2)}{r^2}\right)\psi(r)=0
\end{eqnarray}
Consistency of equations~(\ref{eq4}) and~(\ref{eq5}) demand that 
\begin{equation}
\label{eq6}
\frac{d}{dr}\ln\frac{f^{2}(r)\exp(\int M(g)dg)}{g^{\prime}(r)}=\frac{(D-1)}{r}
\end{equation}
which fixes the form of $f(r)$ as
\begin{equation}
\label{eq7}
f(r)=N r^{\frac{(D-1)}{2}}g^{\prime\frac{1}{2}}\left(\exp\left(-\int M(g)dg\right)\right)^{\frac{1}{2}}
\end{equation}
where $N$ is the integration constant and plays the role of the normalization constant of the wavefunctions.
\\Using (\ref{eq6}) and (\ref{eq7}) in equation (\ref{eq4}) yields
\begin{align}
\label{eq8}
\frac{\psi^{\prime\prime}(r)}{\psi(r)}+\frac{(D-1)}{r}\frac{\psi^{\prime}(r)}{\psi(r)}=-\frac{1}{2}\{g,r\}+\frac{g^{\prime2}(r)}{4}\left[M^{2}(g)+2M^{\prime}(g)-4J(g)\right]-\frac{(D-1)(D-3)}{4r^2}
\end{align}
where the Schwartzian derivative symbol~\cite{Hille}, $\{g,r\}$ is defined as
\begin{equation*}
\{g,r\}=\frac{g^{\prime\prime\prime}(r)}{g^{\prime}(r)}-\frac{3}{2}\frac{g^{\prime\prime2}(r)}{g^{\prime2}(r)}
\end{equation*}
From equations (\ref{eq3}) and (\ref{eq7}), the expression for nomalizable wavefunction is
\begin{equation}
\label{eq9}
\psi(r)=N r^{-\frac{(D-1)}{2}}g^{\prime-\frac{1}{2}}\left(\exp\left(\int M(g)dg\right)\right)^{\frac{1}{2}}Q(g(r))
\end{equation}
The radial wavefunction $\psi(r)=\frac{u(r)}{r}$ has to satisfy the boundary condition $u(r)=0$, in order to rule out singular solutions~\cite{Khelashvili}.\\
Expression (\ref{eq8}) can be cast in the standard Schrodinger equation form (equation (\ref{eq5})) if we can write
\begin{equation}
\label{eq10}
-(E_n-V(r))=-\frac{1}{2}\{g,r\}+\frac{g^{\prime2}(r)}{4}\left[M^{2}(g)+2M^{\prime}(g)-4J(g)\right]-\frac{(D-1)(D-3)}{4r^2}
\end{equation}
Once we choose a particular orthogonal polynomial $Q(g)$ to construct an exact solution of the Schrodinger equation, the characteristic functions of the polynomial $M(g)$, $J(g)$ get specified. We have to choose one or more than one terms containing the function $g(r)$ in expression (\ref{eq10}) and put it equal to a constant to get the energy eigenvalues $E_n$. In our recent paper~\cite{Bhagawati}, we have identified $Q(g)$ as one of the COPs and constructed many new ESPs. In this paper, we identify $Q(g)$ as extended Laguerre polynomial $\hat{L}_n^{\alpha}(x)$ and also as extended Jacobi polynomial $\hat{P}_n^{(\alpha, \beta)}(x)$ and try to construct ESPs associated with them.

\section{Construction of ESPs from exceptional orthogonal polynomials}
\label{sec3}
One of the significant properties of Laguerre or Jacobi type $X_1$ EOPs is that they start with a linear polynomial, unlike the COPs which start with a constant, and still form an orthogonal and complete set with respect to some positive-definite measure. Their properties are given in detail in~\cite{Ullate1, Ullate2} (see Appendix A).

\subsection{Construction of ESPs from Laguerre EOPs}
Identifying
\begin{align}
\label{eq11}
Q(g(r))=\hat{L}_n^{\alpha}(g)
\end{align}
as the exceptional Laguerre polynomial, its characteristic functions $M(g)$ and $J(g)$ are
\begin{align}
\label{eq12}
M(g)=-\frac{(g-\alpha)(g-\alpha+1)}{g(g+\alpha)}
\end{align}
\begin{align}
\label{eq13}
J(g)=\frac{1}{g}\left(\frac{g-\alpha}{g+\alpha}+n-1\right)
\end{align}
Using equations (\ref{eq12}) and (\ref{eq13}) in equation (\ref{eq8}), we obtain
\begin{align}\nonumber
\label{eq14}
\frac{\psi^{\prime\prime}(r)}{\psi(r)}+\frac{(D-1)}{r}\frac{\psi^{\prime}(r)}{\psi(r)}=-\frac{(\alpha^2+2\alpha n-\alpha+2)}{2\alpha}\frac{g^{\prime2}}{g}+\frac{(\alpha+1)(\alpha-1)}{4}\frac{g^{\prime2}}{g^2}+\frac{1}{4}g^{\prime2}\\
+\frac{1}{\alpha}\frac{g^{\prime2}}{(g+\alpha)}+2\frac{g^{\prime2}}{(g+\alpha)^2}-\frac{1}{2}\{g,r\}-\frac{(D-1)(D-3)}{4r^2}
\end{align}
and using equations (\ref{eq11}) and (\ref{eq12}) in (\ref{eq9}) yields
\begin{align}
\label{eq15}
\psi(r)=Nr^{-\frac{(D-1)}{2}}g^{\prime-\frac{1}{2}}\frac{g^{\frac{\alpha+1}{2}}}{(g+\alpha)}\exp{(-\frac{g}{2})}\hat{L}_n^{\alpha}(g(r))
\end{align}
To convert equation (\ref{eq14}) into a standard stationary state Schr\"odinger equation, we make one or more terms of the right hand side of equation (\ref{eq14}) a constant quantity. This enables us to get the energy eigenvalues $E_n$, the functional form of $g(r)$ and subsequently potential $V(r)$ and wavefunction $\psi(r)$.
\\ \\
\hspace*{0.3in}(\romannumeral1) Let us choose
\begin{align}
\label{eq16}
\frac{g^{\prime2}}{g}=p_1^2
\end{align}
where $p_1^2$ is a real positive constant. The functional form of $g(r)$ is
\begin{align}
\label{eq17}
g(r)=\frac{1}{4}p_1^2r^2
\end{align}
Equations (\ref{eq10}), (\ref{eq14}) and (\ref{eq15}) then yield
\begin{align}
\label{eq18}
E_n=\frac{p_1^2}{2}(2n+\alpha-1)
\end{align}
\begin{align}
\label{eq19}
V(r)=\frac{1}{16}p_1^4r^2+\left((\alpha+1)(\alpha-1)+\frac{3}{4}-\frac{(D-1)(D-3)}{4}\right)\frac{1}{r^2}+\frac{4p_1^2}{(p_1^2r^2+4\alpha)}-\frac{32p_1^2\alpha}{(p_1^2r^2+4\alpha)^2}
\end{align}
and
\begin{align}
\label{eq20}
\psi_{n,\ell}(r)=N\frac{r^{\alpha-\frac{(D-2)}{2}}}{(p_1^2r^2+4\alpha)}e^{-\frac{1}{8}p_1^2r^2}\hat{L}_n^{\alpha}(\frac{1}{4}p_1^2r^2)
\end{align}
To get the correct form of centrifugal barrier term in $D$-dimensional Euclidean space, we have to identify the coefficient of $\frac{1}{r^2}$ in potential term (\ref{eq19}) to be $\ell(\ell+D-2)$ \cite{Bhagawati}, which fixes the value of $\alpha$ as
\begin{align}
\label{eq21}
\alpha=\ell+\frac{D-2}{2}
\end{align}
For three dimensional case ($D=3$), let,  $p_1^2=2\omega$ and $n=m+1$. From expressions (\ref{eq18}) and (\ref{eq19}) we get the energy eigenvalues and potential as
\begin{align}
\label{eq22}
E_m=\omega(2m+\ell+\frac{3}{2}) \quad ; \quad m=0,1,2,...
\end{align}
\begin{align}
\label{eq23}
V(r)=V_1(r)+V_2(r)
\end{align}
where
\begin{align}
\label{eq24}
V_1(r)=\frac{1}{4}\omega^2r^2+\frac{\ell(\ell+1)}{r^2}
\end{align}
and
\begin{align}
\label{eq25}
V_2(r)=\frac{4\omega}{(\omega r^2+2\ell+1)}-\frac{8\omega(2\ell+1)}{(\omega r^2+2\ell+1)^2}
\end{align}
for $\omega>0$ and $\ell=0,1,2,...$\\
$V(r)$ is a well-behaved `extended potential' as it incorporates some additional rational terms to the standard radial oscillator potential $V_1(r)$ (see Appendix B). It is clear from equation (\ref{eq22}) that the extended potential has the same energy spectrum as that of the standard one. The corresponding normalized eigenfunctions can be written as
\begin{align}
\label{eq26}
\psi_{m,\ell}(r)=N_m\frac{r^{\ell}}{(\omega r^2+2\ell+1)}e^{-\frac{1}{4}\omega r^2}\hat{L}_{m+1}^{\alpha}(\frac{1}{2}\omega r^2)
\end{align}
where
\begin{align}
\label{eq27}
N_m=\left(\left(\frac{2}{\omega}\right)^{\ell-\frac{3}{2}}\frac{m!}{(m+\ell+\frac{3}{2})\Gamma(m+\ell+\frac{1}{2})}\right)^{\frac{1}{2}}
\end{align}\\
\hspace*{0.3in}(\romannumeral2) Choosing $\frac{g^{\prime2}}{g^2}=p_2^2$, where $p_2^2$ is a real positive constant independent of $r$, we get the functional form of $g(r)$ as
\begin{align}
\label{eq28}
g(r)=e^{-p_2r}
\end{align}
To fulfill the normalizability condition, we consider here only the negative sign in the exponential. Then Equations (\ref{eq10}) and (\ref{eq14}) yield
\begin{align}
\label{eq29}
E_m=-\frac{1}{4}(2A-2m-1)^2p_2^2
\end{align}
where $n=m+1$; \quad $m=0,1,2,...$
\begin{align}
\label{eq30}
V(r)=V_1(r)+V_2(r)-\frac{(D-1)(D-3)}{4r^2}
\end{align}
with
\begin{align}
\label{eq31}
V_1(r)=p_2^2\left(Be^{-p_2r}+\frac{1}{4}e^{-2p_2r}\right)
\end{align}
and
\begin{align}
\label{eq32}
V_2(r)=p_2^2\left((B-A)\frac{e^{-2p_2r}}{(e^{-2p_2r}+\alpha)}+\frac{2e^{-2p_2r}}{(e^{-p_2r}+\alpha)^2}\right)
\end{align}
where $A$ and $B$ are constants and are given as
\begin{align}
\label{eq33}
\frac{\alpha+2n-1}{2}=A \quad ; \quad A+\frac{1}{\alpha}=-B
\end{align}
The function $V_1(r)$ defines a Morse like potential~\cite{Levai1, Dabrowska} (see Appendix B). So, $V(r)$ is an extended Morse potential with the same energy spectrum (\ref{eq29}) as that of Morse potential. It is interesting to note that the potential given by expression (\ref{eq30}) is non-power law and as our formalism suggests, it has an inverse square potential term in spaces where the dimensionality is other than 1 and 3~\cite{Bhagawati}. The wavefunctions of the quantum system can be written as
\begin{align}
\label{eq34}
\psi_{m,\ell=0}(r)=N_mr^{-\frac{(D-1)}{2}}\frac{e^{-\frac{1}{2}\alpha p_2r}}{(e^{-p_2r}+\alpha)}\exp{(-\frac{1}{2}e^{-p_2r})}\hat{L}_{m+1}^{\alpha}(e^{-p_2r})
\end{align}
where normalization constant $N_m$ is given by
\begin{align}
\label{eq35}
N_m=\left(\frac{m!}{(2A-m)\Gamma(2A-m-1)}\right)^{\frac{1}{2}}
\end{align}

\subsection{Construction of ESPs from Jacobi EOPs}
Identifying
\begin{align}
\label{eq36}
Q(g(r))=\hat{P}_n^{(\alpha, \beta)}(g)
\end{align}
as the Jacobi-type $X_1$ poynomial, $n=1,2,3,...$ and  $\alpha, \beta>-1, \quad \alpha \neq \beta$~\cite{Ullate1, Ullate2}. Its characteristic functions are
\begin{align}
\label{eq37}
M(g)=-\frac{(\beta+\alpha+2)g-(\beta-\alpha)}{1-g^2}-\frac{2(\beta-\alpha)}{(\beta-\alpha)g-(\beta+\alpha)}
\end{align}
\begin{align}
\label{eq38}
J(g)=-\frac{(\beta-\alpha)g-(n-1)(n+\beta+\alpha)}{1-g^2}-\frac{(\beta-\alpha)^2}{(\beta-\alpha)g-(\beta+\alpha)}
\end{align}
Using equations (\ref{eq37}) and (\ref{eq38}) in equation (\ref{eq8}), we get
\begin{align}
\label{eq39}\nonumber
\frac{\psi^{\prime\prime}(r)}{\psi(r)}+\frac{(D-1)}{r}\frac{\psi^{\prime}(r)}{\psi(r)}=\frac{(Cg+D_1)g^{\prime2}}{1-g^2}+\frac{(Eg+F)g^{\prime2}}{(1-g^2)^2}+\frac{Gg^{\prime2}}{(\beta-\alpha)g-(\beta+\alpha)}\\
+\frac{Kg^{\prime2}}{[(\beta-\alpha)g-(\beta+\alpha)]^2}-\frac{1}{2}\{g,r\}-\frac{(D-1)(D-3)}{4r^2}
\end{align}
where
\begin{align}\nonumber
C=-\frac{1}{2}\frac{(\beta-\alpha)(\beta+\alpha)}{\beta\alpha}
\end{align}
\begin{align}\nonumber
D_1=-n^2-(\beta+\alpha-1)n-\frac{1}{4}[(\beta+\alpha)^2-2(\beta+\alpha)-4]-\frac{\beta^2+\alpha^2}{2\beta\alpha}
\end{align}
\begin{align}\nonumber
E=-\frac{1}{2}(\beta-\alpha)(\beta+\alpha), \quad \quad F=\frac{1}{2}(\beta^2+\alpha^2-2)
\end{align}
\begin{align}\nonumber
G=-\frac{(\beta-\alpha)^2(\beta+\alpha)}{2\beta\alpha}, \quad \quad K=2(\beta-\alpha)^2
\end{align}
and equation (\ref{eq9}) gives
\begin{align}
\label{eq40}
\psi(r)=Nr^{-\frac{(D-1)}{2}}g^{\prime-\frac{1}{2}}\frac{(1-g)^{\frac{1}{2}(\alpha+1)}(1+g)^{\frac{1}{2}(\beta+1)}}{[(\beta-\alpha)g-(\beta+\alpha)]}\hat{P}_n^{(\alpha, \beta)}(g)
\end{align}
\hspace*{0.3in}(\romannumeral1) Let us choose,
\begin{align}
\label{eq41}
\frac{g^{\prime2}}{1-g^2}=p^2
\end{align}
where $p$ is a constant independent of $r$. We get the functional form of $g(r)$ as
\begin{align}
\label{eq42}
g(r)=\sin{pr}
\end{align}
The parameter $p$ can be set equal to 1 by rescaling the variable $r$. By changing the parameters
\begin{align}\nonumber
\alpha=A-B-\frac{1}{2}, \quad \quad \beta=A+B-\frac{1}{2}
\end{align}
or
\begin{align}\nonumber
A=\frac{1}{2}(\beta+\alpha+1), \quad \quad B=\frac{1}{2}(\beta-\alpha); \quad \quad n=m+1
\end{align}
we get the following results,
\begin{align}
\label{eq43}
E_m=(m+A)^2, \quad \quad m=0,1,2,...
\end{align}
\begin{align}
\label{eq44}
V(r)=V_1(r)+V_2(r)-\frac{(D-1)(D-3)}{4r^2}
\end{align}
with
\begin{align}
\label{eq45}
V_1(r)=[A(A-1)+B^2]\sec^2{r}-B(2A-1)\sec{r}\tan{r}
\end{align}
and
\begin{align}
\label{eq46}
V_2(r)=\frac{2(2A-1)}{(2A-1-2B\sin{r})}-\frac{2[(2A-1)^2-4B^2]}{(2A-1-2B\sin{r})^2}
\end{align}
$V_1(r)$ can be identified as Scarf I potential (see Appendix B). So, $V(r)$ is an extended potential with the same behavior as $V_1(r)$. Its wavefunctions  can be written as
\begin{align}
\label{eq47}
\psi_{m,\ell=0}(r)=N_mr^{-\frac{(D-1)}{2}}\frac{(1-\sin{r})^{\frac{1}{2}(A-B)}(1+\sin{r})^{\frac{1}{2}(A+B)}}{(2A-1-2B\sin{r})}\hat{P}_{m+1}^{(A-B-\frac{1}{2}, A+B-\frac{1}{2})}(\sin{r})
\end{align}
where
\begin{align}
\label{eq48}
N_m=\left(\frac{m!(2m+2A)\Gamma(m+2A)}{2^{2A-2}(m+A-B+\frac{1}{2})(m+A+B+\frac{1}{2})\Gamma(m+A-B-\frac{1}{2})\Gamma(m+A+B-\frac{1}{2})}\right)^{\frac{1}{2}}
\end{align}\\

\hspace*{0.3in}(\romannumeral2) By assuming,
\begin{align}
\label{eq49}
\frac{g^{\prime2}}{(1-g^2)^2}=c^2
\end{align}
where $c^2$ is a real positive constant, we get
\begin{align}
\label{eq50}
g(r)=\tanh{cr}
\end{align}
Taking $c=1$ and rearranging some parameters
\begin{align}\nonumber
m+A=P_1, \quad \quad B(2P_1-2m-1)=Q
\end{align}
we arrive at the following results
\begin{align}
\label{eq51}
E_m=-(P_1-m-\frac{1}{2})^2-\frac{Q^2/4}{(P_1-m-\frac{1}{2})^2}, \quad \quad m=0,1,2,...
\end{align}
\begin{align}
\label{eq52}
V(r)=V_1(r)+V_2(r)
\end{align}
with
\begin{align}
\label{eq53}
V_1(r)=-(P_1^2-\frac{5}{4})\sech^2{r}-Q\tanh{r}
\end{align}
\begin{align}
\label{eq54}
V_2(r)=\frac{4(2-\tanh^2{r})}{(2\sinh{r}-Q_1\cosh{r})^2}
\end{align}
where $Q_1=\frac{(2P_1-2m-1)^2}{Q}$. By looking at the energy spectrum and the form of the potential $V_1(r)$ (see Appendix B), we can conclude that the overall potential $V(r)$ is an extended Rosen-Morse like potential. Its normalized wavefunctions can be written as
\begin{align}
\label{eq55}
\psi_{m,\ell=0}(r)=N_mr^{-\frac{(D-1)}{2}}(\cosh{r})^{\frac{1}{2}}\frac{(1-\tanh{r})^{\frac{\lambda}{2}}(1+\tanh{r})^{\frac{\delta}{2}}}{[\frac{2Q}{(2P_1-2m-1)}\tanh{r}-(2P_1-2m-1)]}\hat{P}_{m+1}^{(\lambda-\frac{1}{2},\delta-\frac{1}{2})}(\tanh{r})
\end{align}
where
\begin{align}\nonumber
\lambda=P_1-m-\frac{Q}{2P_1-2m-1}, \quad \quad \delta=P_1-m+\frac{Q}{2P_1-2m-1}
\end{align}
The normalization constant $N_m$ is given by
\begin{align}
\label{eq56}
N_m=\left[\frac{2P_1m!\Gamma(2P_1-m)}{2^{2P_1-2m-2}(m+\lambda+\frac{1}{2})(m+\delta+\frac{1}{2})\Gamma(m+\lambda-\frac{1}{2})\Gamma(m+\delta-\frac{1}{2})}\right]^{\frac{1}{2}}
\end{align}

\section{Discussion and Conclusion}
\label{sec4}
In this paper, we have constructed four new exactly solvable quantum mechanical potentials which give rise to bound state solutions of $D$-dimensional radial Schr\"odinger equation. To achieve this, we employed a simple transformation method which comprises a co-ordinate transformation followed by a functional transformation. Laguerre or Jacobi type $X_1$ exceptional orthogonal polynomials play an important role in constructing extended exactly solvable potentials. While the extended radial oscillator potential and the extended Scarf I potential have already been introduced in~\cite{Quesne1}, in this paper we have re-derived these potentials in $D$-dimensional Euclidean space. The extended Morse potential and the extended Rosen-Morse potential are completely new and have not been covered in the literature so far.

\appendix
\numberwithin{equation}{section}

\section{Appendix}

\subsection{Basic properties of Laguerre EOPs}
The $X_1$ Laguerre EOP, denoted by $\hat{L}_n^{\alpha}(z)$; $n=1,2,3,...$, $\alpha>0$, has following properties~\cite{Ullate1}:
\begin{align}
\label{57}
\hat{L}_1^{\alpha}(z)=-z-\alpha-1,\quad\quad \hat{L}_2^{\alpha}(z)=z^2-\alpha(\alpha+2),...
\end{align}
\begin{align}
\label{58}
\left(z\frac{d^2}{dz^2}-\frac{z-\alpha}{z+\alpha}\left[(z+\alpha+1)\frac{d}{dz}-1\right]\right)\hat{L}_n^{\alpha}(z)=-(n-1)\hat{L}_n^{\alpha}(z)
\end{align}
\begin{align}
\label{59}
\int_0^{\infty}\hat{L}_{n^{\prime}}^{\alpha}(z)\hat{L}_n^{\alpha}(z)\frac{z^{\alpha}e^{-z}}{(z+\alpha)^2}dz=\delta_{n^{\prime},n}\frac{\Gamma(n+\alpha+1)}{(n+\alpha-1)(n-1)!}
\end{align}
\begin{align}
\label{60}
\hat{L}_n^{\alpha}(z)=nL_n^{\alpha}-2(n+\alpha)L_{n-1}^{\alpha}(z)+(n+\alpha)L_{n-2}^{\alpha}(z)
\end{align}

\subsection{Basic properties of Jacobi EOPs}
The $X_1$ Jacobi EOP, denoted by  $\hat{P}_n^{(\alpha, \beta)}(z)$; $n=1,2,3,...$, $\alpha, \beta>-1, \quad \alpha \neq \beta$, has following properties~\cite{Ullate1}:
\begin{align}
\label{61}
\hat{P}_1^{(\alpha,\beta)}(z)=-\frac{1}{2}z-\frac{2+\alpha+\beta}{2(\alpha-\beta)},\quad\quad
\hat{P}_2^{(\alpha,\beta)}(z)=-\frac{\alpha+\beta+2}{4}z^2-\frac{\alpha^2+\beta^2+2(\alpha+\beta)}{2(\alpha-\beta)}z-\frac{\alpha+\beta+2}{4},...
\end{align}
\begin{align}
\label{62}
\left[(z^2-1)\frac{d^2}{dz^2}+2a\left(\frac{1-bz}{b-z}\right)\left((z-c)\frac{d}{dz}-1\right)\right]\hat{P}_n^{(\alpha,\beta)}(z)=(n-1)(\alpha+\beta+n)\hat{P}_n^{(\alpha,\beta)}(z)
\end{align}
where the real parameters $a$, $b$ and $c$ are given by
\begin{align}\nonumber
a=\frac{\beta-\alpha}{2}; \quad\quad b=\frac{\beta+\alpha}{\beta-\alpha}; \quad\quad c=b+\frac{1}{a}
\end{align}
\begin{align}
\label{63}
\int_{-1}^1\frac{(1-z)^{\alpha}(1+z)^{\beta}}{(z-b)^2}\left(\hat{P}_n^{(\alpha,\beta)}(z)\right)^2dz=\frac{(\alpha+n)(\beta+n)}{4(\alpha+n-1)(\beta+n-1)}C_{n-1}
\end{align}
where
\begin{align}\nonumber
C_n=\frac{2^{\alpha+\beta+1}}{(\alpha+\beta+2n+1)}\frac{\Gamma(\alpha+n+1)\Gamma(\beta+n+1)}{\Gamma(n+1)\Gamma(\alpha+\beta+n+1)}
\end{align}
\begin{align}
\label{64}
\hat{P}_n^{(\alpha,\beta)}(z)=-f_nP_n^{(\alpha,\beta)}(z)+2b g_n P_{n-1}^{(\alpha,\beta)}(z)-h_n P_{n-2}^{(\alpha,\beta)}(z)
\end{align}
where
\begin{align}\nonumber
f_n=\frac{n(\alpha+\beta+n)}{(\alpha+\beta+2n-1)(\alpha+\beta+2n)}
\end{align}
\begin{align}\nonumber
g_n=\frac{(\alpha+n)(\beta+n)}{(\alpha+\beta+2n-2)(\alpha+\beta+2n)}
\end{align}
\begin{align}\nonumber
h_n=\frac{(\alpha+n)(\beta+n)}{(\alpha+\beta+2n-2)(\alpha+\beta+2n-1)}
\end{align}

\section{Appendix}

\subsection{Standard ESPs referred}

The radial oscillator potential ($V_{RO}$) in $D$-dimensional Euclidean spcae~\cite{Bhagawati} is
\begin{align}
\label{65}
V_{RO}(r)=\frac{1}{4}\omega^2r^2+\frac{\ell(\ell+D-2)}{r^2}
\end{align}
The corresponding exact energy eigenvalues and the wavefunction are given by
\begin{align}
\label{66}
E_{n_r}^{RO}=\omega(2n_r+\ell+\frac{D}{2})
\end{align}
\begin{align}
\label{66}
\psi_{RO}(r)=Nr^\ell\exp(-\frac{\omega r^2}{4})L_{n_r}^{\ell+\frac{D-2}{2}}(\frac{\omega r^2}{2})
\end{align}
The Morse potential ($V_M$) is given by~\cite{Levai1, Dabrowska}
\begin{align}
\label{67}
V_M(r)=-B(2A+a)e^{-ar}+B^2e^{-2ar}
\end{align}
The corresponding exact energy eigenvalues and the wavefunction are given by
\begin{align}
\label{68}
E_n^M=-(A-na)^2
\end{align}
\begin{align}
\label{69}
\psi_M(r)=Ng^{s-n}e^{-\frac{g}{2}}L_n^{(2s-2n)}(g(r))
\end{align}
where
\begin{align}\nonumber
g(r)=\frac{2B}{a}e^{-ar}; \quad\quad s=\frac{A}{a}
\end{align}
The Scarf I potential ($V_S$) is given by~\cite{Castillo}
\begin{align}
\label{70}
V_S(r)=(a^2+b^2-a\alpha)\sec^2{\alpha r}-b(2a+\alpha)\tan{\alpha r}\sec{\alpha r}
\end{align}
The corresponding exact energy eigenvalues and the wavefunction are given by
\begin{align}
\label{71}
E_n^S=(n\alpha+a)^2
\end{align}
\begin{align}
\label{72}
\psi_S(r)=N(1-\sin{\alpha r})^\frac{\gamma}{2}(1+\sin{\alpha r})^\frac{\delta}{2}P_n^{\gamma-\frac{1}{2},\delta-\frac{1}{2}}(\sin{\alpha r})
\end{align}
where
\begin{align}\nonumber
\gamma=\frac{a-b}{\alpha}; \quad\quad \delta=\frac{a+b}{\alpha}
\end{align}
The Rosen-Morse potential ($V_{RM}$) is given by~\cite{Dabrowska}
\begin{align}
\label{73}
V_{RM}(r)=-A(A+a)\sech^2{ar}+2B\tanh{ar}
\end{align}
The corresponding exact energy eigenvalues and the wavefunction are given by
\begin{align}
\label{74}
E_n^{RM}=-(A-na)^2-\frac{B^2}{(A-na)^2}
\end{align}
\begin{align}
\label{75}
\psi_{RM}(r)=N(1-\tanh{ar})^{\frac{s-n+a}{2}}(1+\tanh{ar})^{\frac{s-n-a}{2}}P_n^{(s-n+a,s-n-a)}(\tanh{ar})
\end{align}
where
\begin{align}\nonumber
s=\frac{A}{a}
\end{align}

\section*{Acknowledgements}
The author is indebted to Prof. S. A. S. Ahmed for his valuable suggestions on the subject and thanks the UGC-RFSMS, India for financial support.

\end{document}